\documentclass[nofootinbib,
tightenlines,
superscriptaddress,11pt]{revtex4}

\usepackage{graphicx}
\usepackage{amsmath,amssymb,amsfonts,amsthm,stmaryrd,mathtools,bm,physics}
\usepackage{xcolor}
\usepackage{tikz}
\usepackage[normalem]{ulem}
\usepackage{braket}
\allowdisplaybreaks[1]

\def\be#1\ee{\begin{align}#1\end{align}}

\def\ba{\begin{eqnarray}}
\def\ea{\end{eqnarray}}

\usepackage[bookmarks,linktocpage, colorlinks=true, plainpages = false, citecolor = treegreen,  linkcolor=darkblue, urlcolor = darkblue, filecolor = blue]{hyperref} 

\definecolor{darkblue}{rgb}{0., 0.4, 0.8}
\definecolor{treegreen}{rgb}{0., 0.7, 0.3}

\begin{document}

\title{Lower bound of black hole hair in pure Lovelock theory of gravity}

\author{Pabitra Tripathy}
\email{pabitra.tripathy@saha.ac.in} 
\affiliation{Theory Division, Saha Institute of Nuclear Physics, 
		1/AF, Bidhannagar, Kolkata 700064, India}
\affiliation{Homi Bhabha National Institute, Anushaktinagar, 
		Mumbai, Maharashtra 400094, India}

\begin{abstract}
As an alternative to the ``no hair conjecture," the ``no short hair conjecture" for hairy black holes was established earlier. This theorem stipulates that hair must be present above 3/2 of the event horizon radius for a hairy black hole. It is assumed that the nonlinear behavior of the matter field plays a key role in the presence of such hair. Subsequently, it was established that the hair must extend beyond the photon sphere of the corresponding black hole. We have investigated the validity of the ``no short hair conjecture" in pure Lovelock gravity. Our analysis has shown that irrespective of dimensionality and Lovelock order, the hair of a static, spherically symmetric black hole extends at least up to the photon sphere. 

\end{abstract}


\maketitle
\tableofcontents


\newpage
\section{Introduction}\label{Sec:Introduction}
Despite their complex interior configurations, black holes can be fully characterized by a few conserved charges measured at asymptotic infinity, specifically, mass, electric charge, magnetic charge, and angular momentum. This concept is central to the ``no hair conjecture (NHC)"  \cite{John A. Wheeler}. According to the NHC, a matter field without conserved charge at asymptotic infinity cannot be stabilized in the presence of a black hole; it will either radiate away or be absorbed by the black hole \cite{Nunez:1996xv}. As a result, the NHC excludes the presence of scalar \cite{J. E. Chase}, massive vector \cite{VECTOR}, and spinor \cite{spinor} fields from the exterior of a stationary black hole.

However, this idea was challenged as various ``hairy" black holes were discovered, requiring additional parameters unrelated to conserved charges for complete specification (referred to as ``hair"). Examples include ``colored black holes," which are solutions of the static SU(2) Einstein-Yang-Mills (EYM) equation and possess additional Yang-Mills hair \cite{color}. Other examples include black hole solutions in Einstein-Skyrme, Einstein-Yang-Mills-Dilaton, Einstein-Yang-Mills-Higgs, and Einstein-Nonabelian-Proca theories \cite{sky, dilaton, Higgs, EYM, proca}. The stability of these black holes has been studied, and not all of them are unstable \cite{inst1, inst2, inst3, inst4}. Another counterexample to the uniqueness theorem is based on multi-horizon configurations in asymptotically de Sitter spacetime, where two black holes can be in static equilibrium due to a balance between gravitational pull and cosmic expansion, violating the uniqueness theorem without matter \cite{desitter}.

This existence of hair challenges the NHC's validity and raises questions about the assumed behavior of the field that led to its establishment. It has become clear that the interaction between matter content near the horizon and in the faraway region is a key factor in the existence of such hair \cite{Nunez:1996xv}. This raises the question of whether a lower bound of matter content following nontrivial character is necessary, or in other words, what is the minimum length of hair, and can it become arbitrarily small? The ``no short hair conjecture" elegantly demonstrates a universal lower bound on the length of hair ($r_{\text{hair}}$) \cite{Nunez:1996xv}, stating that the nonlinear behavior of the matter field extends beyond 3/2 of the horizon radius ($r_H$), terming this region the ``hairosphere."

Null circular geodesics play a crucial role in understanding the geometric structure of black hole spacetime, with the photon orbit separating the exterior of the black hole spacetime into two regions, below which no test particle can orbit the black hole. The photon orbit of Schwarzschild spacetime is $3 r_H/2 $. This connection with the hairosphere in the Schwarzschild case naturally raises the question: is there a connection between the photon orbit and the lower bound of the nontrivial behavior of the near-horizon field? A fascinating development showed that the hair must extend at least up to the height of the photon sphere ($r_{\text{photon}}$) of a static spherically symmetric black hole \cite{photonsphere}. Specifically, it was shown that $r_{\text{hair}} \geq r_{\text{photon}}$. This finding increases the importance of the no short hair conjecture from an observational perspective, as it implies that if hair is present, it must extend at least up to the photon orbit, making the study of the near-photon orbit region sufficient for detection. The gradual improvement in the precision of gravitational wave detection and analysis has enabled more accurate probing of the near-field region, offering the possibility of testing the signature of black hole hair \cite{eht, eht1, eht2, eht3, DDIMENSION}.

It is important to note that the formal proof of the no short hair conjecture and its connection with the photon orbit is limited to Einstein's theory and four dimensions. However, it was shown that the no short hair theorem can be generalized to arbitrary spacetime dimensions ($d \geq 4$) and other theories of gravity, stating that regardless of the theory of gravitation and dimension, the hair of any static, spherically symmetric, asymptotically flat black hole must extend at least up to the innermost light ring \cite{DDIMENSION}. This generalization aids in studying the no short hair theorem in modified theories of gravity and dismisses the possibility of using it to describe the theory.

Motivated by this generalization of the no short hair conjecture in arbitrary dimensions, this study investigates the theorem for pure Lovelock gravity without restrictions in dimension and Lovelock order. We verify that for a static, spherically symmetric, asymptotically flat pure Lovelock black hole, the nontrivial characteristic of the matter field must extend at least up to the photon orbit. Pure Lovelock gravity is a theory distinguished by a modified kinetic term in dimensions greater than 4, expressed through a polynomial in the Riemann curvature of order N. When N is set to 1, the theory simplifies to the linear term, corresponding to scalar curvature and aligning with Einstein's General Relativity (GR). In the case of four dimensions, N is restricted to 1. Generally, the values of d and N are independent, except for the constraint that N greater than one is only allowed in dimensions greater than 4, while it consistently holds that N is less than [(d-1)/2]. Despite the presence of higher curvature terms in the Lagrangian for pure Lovelock theories, the field equations remain second-order \cite{Gannouji:2019gnb, Dadhich:2015nua, Dadhich:2012cv, Gannouji:2013eka, Dadhich:2015ivt, Padmanabhan:2013xyr, Dadhich:2015lra}.

The structure of this paper unfolds as follows: In Section \ref{lowerbound}, we delve into the computation of the lower bound of black hole hair within this theory. Our conclusions, detailed in Section \ref{discussion}, encapsulate a thorough discussion of our discoveries.

\emph{Notations and Conventions:} Throughout our calculations, we adopt the convention $c=\hbar=G=1$. Additionally, our metric adheres to the mostly positive signature convention, where in four spacetime dimensions, the Minkowski metric is represented by $\eta_{\mu \nu}=\textrm{diag.}(-1,+1,+1,+1)$.

\section{Mechanisms Supporting Scalar Clouds Around Black Holes}

The existence of hairy black holes serves as the cornerstone of the ``no short hair" conjecture, which posits that the interaction between near-horizon fields and fields at far distances is essential for the formation of black hole hair \cite{Nunez:1996xv}. Although a comprehensive mechanism remains elusive, several recent studies have provided insights into the conditions leading to stable cloud configurations outside black holes \cite{Hod_2012, Hod_2013, Barranco_2012, Herdeiro_2014, Hod_2014}. One such study examines the formation of stationary charged scalar clouds around Kerr-Newman black holes \cite{Hod_2014}. In this section, we briefly revisit this mechanism. For a detailed discussion, we refer the reader to Ref. \cite{Hod_2014} and references therein.

The formation of a stationary charged scalar cloud can be attributed to two distinct physical mechanisms: superradiant scattering of the scalar field and the gravitational attraction between the black hole and the massive scalar field \cite{Hod_2014}.

 \begin{itemize}
    \item Superradiance is a remarkable physical process in which waves interacting with a rotating or charged black hole are amplified, extracting energy and angular momentum from the black hole. When a wave interacts with a black hole, part of the wave may be absorbed, while another part is scattered. Under specific conditions, the scattered wave emerges with more energy than it initially had, drawing this additional energy from the black hole's rotational energy or electromagnetic field.

The amplification occurs when the wave satisfies the superradiance condition:
\begin{equation}
    \omega<m\Omega_H+q\Phi_H,
\end{equation}
Where, $\omega, m,  q, $ are the frequency, azimuthal harmonic number, and charge of the wave. $\Omega_H, \Phi_H$ represents the black hole's angular velocity and electric potential respectively. 

This energy exchange mechanism ensures that certain wave modes are amplified instead of being completely absorbed or dissipated, allowing the scalar field to sustain itself near the black hole.
\end{itemize}

\begin{itemize}
    \item The second mechanism underlying the formation of scalar clouds is the gravitational attraction between the massive scalar field and the black hole. The mass $\mu$ of the scalar field introduces a natural trapping mechanism by acting as a potential barrier that confines specific field modes near the black hole.

In the Klein-Gordon equation governing the scalar field, the mass term creates a ``reflective barrier" for wave modes with frequencies below the mass threshold:
\begin{equation}
    \omega^2<\mu^2.
\end{equation}
In this regime, the wave modes lack sufficient energy to escape to spatial infinity, leading to their confinement near the black hole. This behavior is analogous to a particle being trapped in a potential well. The scalar field’s mass thus plays a crucial role in creating an effective potential that prevents low-frequency modes from dispersing.
\end{itemize}

The interplay between superradiance and the gravitational attraction due to the scalar field's mass underpins the trapping of scalar fields around rotating black holes. These mechanisms work together as follows.

Superradiance amplifies specific wave modes by extracting energy and angular momentum from the black hole, preventing the scalar field from being entirely absorbed into the black hole.

The scalar field’s mass creates an effective potential barrier at large distances, confining low-frequency modes in the black hole’s vicinity.

Gravitational attraction further localizes the field, ensuring it remains spatially confined.

The synergy between these effects balances the dynamics of energy and angular momentum, enabling the formation of stationary or long-lived scalar field configurations, commonly referred to as scalar clouds, around the black hole.

  \subsection{Analtyical description}
  We consider a system with a charged scalar field $\Phi$ coupled to a Kerr-Newman black hole \cite{Hod_2014}. The line element of the Kerr-Newman spacetime in Boyer-Lindquist coordinate $(t,r,\theta, \phi)$ is given by
  \begin{equation}
      ds^2=-\frac{\Delta}{\rho^2}(dt-a \sin^2{\theta}d\phi)^2+\frac{\rho^2}{\Delta}dr^2+\rho^2d\theta^2+\frac{\sin^2{\theta}}{\rho^2}[adt-(r^2+a^2)d\phi]^2,
  \end{equation}
where $\Delta\equiv r^2-2Mr+a^2+Q^2$ and $\rho\equiv r^2+a^2\cos^2{\theta}$. $M, aM \,\,\text{and}\,\, Q$ are the black hole’s mass, angular momentum, and charge, respectively.

The Klein-Gordon equation of the charged massive scalar field $\Phi(t,r,\theta,\phi)$ gives the dynamics of the field in the background of the Kerr-Newmann black hole, is given by
\begin{equation}
    [(\nabla^\mu-iqA^\mu)(\nabla_\mu-iqA_\mu)-m_f^2]\Phi=0,
\end{equation}
where $m_f, q$ are the mass and charge coupling of the field and $A_\mu$ is the electromagnetic potential of the charged black hole. 

The field $\Phi$ can be expressed as the sum of modes
\begin{equation}
  \Phi(t,r,\theta,\phi)  =\sum_{l,m}e^{-i\omega t}e^{im\phi}R_{lm}(r)\Theta_{lm}(\theta)
\end{equation}
Substituting this ansatz into the Klein-Gordon equation and applying the separation of variables yields two decoupled equations: one for the radial function $R_{lm}$ and one for the angular function $\Theta_{lm}$. The solutions of the angular equation are the spheroidal harmonics. 

The angular equation solutions are spheroidal harmonics, while the radial equation is the radial Teukolsky equation,
\begin{equation}
    \frac{d}{dr}\left(\Delta\frac{dR_{lm}}{dr}\right)+\left[\frac{((r^2+a^2)\omega-am-qQr)^2}{\Delta}+2ma\omega-m_f^2(r^2+a^2)-A\right]R_{lm}=0.
\end{equation}
\newline \textit{Boundary conditions:}
The physical behavior of the scalar field is determined by the boundary conditions at spatial infinity and at the black hole horizon.
\begin{itemize}
    \item The field must decay exponentially, ensuring it does not radiate away. The asymptotic behavior is,
    \begin{equation}\label{boundary1}
        R_{lm}(r\rightarrow\infty)\approx\frac{1}{r}e^{-\sqrt{m_f^2-\omega^2}r}.
    \end{equation}
    which imposes the restriction,
\begin{equation}
    \omega^2<m_f^2
\end{equation}
This inequality reflects that low-frequency modes lack sufficient energy to overcome the potential barrier created by the scalar field’s mass $m_f$. The mass term acts as a natural confining mechanism, preventing the field from ``spreading out" to infinity.

    \item Near the event horizon: At the black hole's outer horizon the field behaves as a purely ingoing wave,
    \begin{equation}\label{boundary2}
        R_{lm}(r\rightarrow r_+)\approx e^{-i(\omega-\omega_c)r_*},
    \end{equation}
    where $r_*$ is the `tortoise' coordinate defined by $\frac{dr_*}{dr}=\frac{r^2+a^2}{\Delta}$ and $\omega_c=m\Omega_H+q\Phi_H$ is the critical superradiant frequency.

    This boundary condition ensures no field escapes from the interior of the black hole. Additionally, it requires $Im(\omega)=0$, maintaining the oscillatory nature of the wave.
\end{itemize}
The combination of these boundary conditions imposes quantization on the allowed frequencies $\omega$, leading to a discrete spectrum of resonant frequencies. These frequencies depend on the black hole parameters $(M, Q, a)$,  the field parameters $(q,m_l)$, and the mode numbers $(n,l,m)$. Only specific frequencies $\omega$ satisfy the boundary conditions at both the horizon and spatial infinity, resulting in stationary configurations of the scalar field.

The boundary conditions and the radial equation together describe a scalar field that is confined in the region between the event horizon and spatial infinity.

Horizon Constraint: The ingoing boundary condition ensures that no field escapes the black hole interior, preventing accumulation at the horizon.

Infinity Constraint: The exponential decay ensures the field is localized and does not propagate to infinity.

Intermediate Region: In the region between the horizon and infinity, the field oscillates, forming a stable configuration trapped by the effective potential well.
The effective potential governs the field’s behavior, acting as a ``barrier" that confines low-frequency modes near the black hole. The resulting configuration represents a stationary bound state—a long-lived scalar cloud around the black hole.

\section{Do Bound States Exist for a Pure Lovelock Black Hole in Higher Dimensions?}

In this article, we have shown that if hair exists in a pure Lovelock black hole, then the lower bound on the length of the hair corresponds to the photon orbit. However, it is well known that in general relativity, no bound orbits exist for black holes in spacetime dimensions \( D > 4 \). This raises a critical question: does a pure Lovelock black hole in \( D > 4 \) permit bound orbits at all? This issue has been discussed extensively in Ref.~\cite{Dadhich_2013}. Here, we revisit this important topic to underscore its relevance in the context of our findings.

To establish the existence of bound orbits, the effective potential \( V(r) \) must possess a minimum. This requires,
\begin{equation}\label{minimum condition}
    V'(r) = 0, \quad V''(r) > 0,
\end{equation}
where a prime denotes the derivative with respect to \( r \). For a system governed by a central attractive force, the effective potential takes the form
\begin{equation}
    V(r) = -\frac{m}{r^{d-2}} + \frac{l^2}{r^2},
\end{equation}
where \( m \), \( l \), and \( d \) represent the mass, angular momentum, and spatial dimensionality, respectively. The condition for bound orbits imposes the following restriction on \( d \),
\begin{equation}
    d - 2 < 2.
\end{equation}
This constraint implies that for systems whose dynamics are governed by the Laplace equation, bound orbits can occur only in three spatial dimensions (\( d = 3 \)).

Analogous to Newtonian gravity, general relativity also forbids bound orbits in spacetime dimensions greater than \( D = 4 \). The line element of a static, spherically symmetric spacetime is given by:
\begin{equation}
    ds^2 = -f(r) dt^2 + f(r)^{-1} dr^2 + r^2 d\Omega^2_{D-2},
\end{equation}
where \( d\Omega^2_{D-2} \) represents the metric on a \( (D-2) \)-dimensional sphere.

For this spacetime, the effective potential can be expressed as:
\begin{equation}
    V(r) = f(r) \left(\frac{l^2}{r^2} + 1\right).
\end{equation}
The conditions for bound orbits, given in Eq.~(\ref{minimum condition}), then lead to:
\begin{equation}\label{eq1a}
    \frac{l^2}{r^2} = \frac{r f'}{2f - r f'},
\end{equation}
and
\begin{equation}\label{eq1b}
    r f f'' + 3f f' - 2r f'^2 > 0.
\end{equation}
For a general \( D \)-dimensional static black hole where the metric function \( f(r) \) satisfies the Laplace equation, we assume:
\begin{equation}\label{eq1c}
    f(r) = 1 - \frac{M}{r^{D-3}},
\end{equation}
where \( M \) is the mass parameter of the black hole. Using Eqs.~(\ref{eq1a}), (\ref{eq1b}), and (\ref{eq1c}), the lower bound for the radius of circular orbits is obtained as:
\begin{equation}
    r_0 > r_{\text{ph}} = M^{1/(D-3)} \left(\frac{D-1}{2}\right)^{1/(D-3)},
\end{equation}
where \( r_{\text{ph}} \) is the radius of the photon orbit. The radius of the innermost stable circular orbit (ISCO) is then given by:
\begin{equation}
    r_{\text{ISCO}} = M^{1/(D-3)} \left(\frac{D-1}{5-D}\right)^{1/(D-3)}.
\end{equation}
It follows that in general relativity, bound orbits cannot exist in spacetime dimensions greater than \( D = 4 \), as the stability conditions fail to hold.

Pure Lovelock gravity is a generalization of Einstein gravity to higher dimensions, where the action includes only a single polynomial term of degree \( N \) in the curvature tensor. In odd-dimensional spacetimes (\( D = 2N+1 \)), the equations of motion for pure Lovelock gravity lead to a spacetime curvature that is locally flat. This implies that the gravitational field equations do not support propagating degrees of freedom, and the spacetime geometry is entirely determined by the matter distribution, with no ``dynamical" interplay \cite{Dadhich_2012a}. Conversely, in even-dimensional spacetimes (\( D = 2N+2 \)), the equations of motion permit non-trivial curvature, allowing the gravitational field to respond dynamically to the presence of matter and energy.

For pure Lovelock gravity in even spacetime dimensions, the effective potential is given by \cite{Dadhich_2013}:
\begin{equation}
    V(r) = \left[1 - \left(\frac{M}{r}\right)^{1/N}\right] \left(\frac{l^2}{r^2} + 1\right),
\end{equation}
where \( M \) is the mass parameter, \( l \) is the angular momentum, and \( N \) is the order of the Lovelock polynomial. It is important to emphasize that in this case, the force governing the system is \textit{not Laplace-driven}.

Analyzing the minimum of the effective potential, the minimum radius for the existence of circular orbits is given by:
\begin{equation}
    r_0 > r_{\text{ph}} = M \left(\frac{2N+1}{2N}\right)^N,
\end{equation}
where \( r_{\text{ph}} \) represents the radius of the photon orbit. From a stability analysis, the radius of the innermost stable circular orbit (ISCO) is determined as:
\begin{equation}
    r_{\text{ISCO}} = M \left(\frac{2N+1}{2N-1}\right)^N.
\end{equation}

By substituting \( N=1 \), these expressions reduce to the well-known results for the \( 4D \) Schwarzschild case:
\[
r_{\text{ph}} = \frac{3}{2}M \quad \text{and} \quad r_{\text{ISCO}} = 3M.
\]

Thus, in pure Lovelock gravity, bound orbits always exist in all even-dimensional spacetimes (\( D = 2N+2 \)). The key reason behind this unique feature lies in the nature of the force, which is not Laplace-driven. This distinction is a remarkable feature of pure Lovelock gravity, setting it apart from Einstein gravity and other generalizations.

\color{black}


\section{Bound in black hole hair in pure Lovelock theory} \label{lowerbound}
We consider static, spherically symmetric, asymptotically flat black-hole spacetime in pure Lovelock gravity. The line element can be written in the following form,
\begin{equation}\label{metric}
    ds^2=-e^\nu dt^2+e^\lambda dr^2+r^2d\Omega^2_{d-2}~,
\end{equation}
where $\nu$ and $\lambda$ are functions of radial coordinate $r$ and $d\Omega^2_{d-2}$ is the line element on the surface of a $d-2$ dimensional sphere.

Asymptotic flatness requires that as $r\rightarrow \infty$,
\begin{equation}
    e^\nu\rightarrow 1\,\,\,\,\,\,\,\text{and}\,\,\,\,\,\, e^\lambda\rightarrow 1~,
\end{equation}
At the event horizon $r=r_H$,
\begin{equation}\label{lambdarh}
    e^{-\lambda(r_H)}=0~.
\end{equation}

We take the energy-momentum tensor components as $T^t_t=-\rho$, $T^r_r=T^\theta_\theta=T^\psi_\psi=...=p$, where $\rho$ is the energy density and $p$ is the radial pressure. Without loss of generality, we have taken all tangential pressure equal to the radial pressure to keep the analytical expression simple.

The corresponding field equations can be written as \cite{naresh},
\begin{equation}\label{rho}
    8\pi \rho(r)=\frac{(1-e^{-\lambda})^{N-1}}{2^{N-1}r^{2N}}[rN\lambda'e^{-\lambda}+(d-2N-1)(1-e^{-\lambda})]
\end{equation}
and
\begin{equation}\label{p}
    8\pi p(r)=\frac{(1-e^{-\lambda})^{N-1}}{2^{N-1}r^{2N}}[rN\nu'e^{-\lambda}-(d-2N-1)(1-e^{-\lambda})]~.
\end{equation}
From the conservation of energy-momentum tensor, i.e., $\nabla_\mu T^{\mu \nu}=0$ we get,
\begin{equation}\label{conservation}
    2p'=-\nu' (\rho+p)~.
\end{equation}
Here, all prime denotes the derivative with respect to the radial coordinate $r$.

Substituting Eq. \ref{p} in Eq. \ref{conservation} we get the pressure gradient as follows,
\begin{equation}\label{pprime}
    p'(r)=-\frac{(\rho+p)}{2}\frac{2^{N+2}\pi p r^{2N}+(d-2N-1)(1-e^{-\lambda})^N}{Nre^{-\lambda}(1-e^{-\lambda})^{N-1}}~.
\end{equation}
Now we consider a function $P(r)=r^dp(r)$. We analyze the behavior of this function in the horizon and near-horizon regions. $P(r)$ represents the behavior of the matter field in the black hole spacetime. Our goal is to find out the region where $P(r)$ depicts a nontrivial behavior, i.e., $P(r)\neq 0$, and we follow the prescription of the paper \cite{photonsphere}. 

Using Eq. \ref{pprime}, we find the gradient of the function $P$ as follows,
\begin{equation}\label{Pprime}
\begin{aligned}
    P'(r)=&d~r^{d-1}p\\&-\frac{(\rho+p)}{2}\frac{2^{N+2}\pi p r^{2N}+(d-2N-1)(1-e^{-\lambda})^N}{Nr^{1-d}e^{-\lambda}(1-e^{-\lambda})^{N-1}}~.
\end{aligned}
\end{equation}
We assume some conditions on the energy-momentum tensor of the matter field outside the horizon. The conditions are the following:
\begin{enumerate}
    \item The weak energy condition must be satisfied, which implies,
    \begin{equation}
        \rho\geq0,\,\,\,\,\,\,\,\, \text{and}\,\,\,\,\,\,\,\rho+p\geq0~.
    \end{equation}

    \item The trace of energy-momentum tensor is non-positive, i.e.,
    \begin{equation}\label{trace}
        T\leq0~,
    \end{equation}
    which implies, $\rho\geq(d-1)p$~.

    \item  There is no conserved charge at asymptotic infinity associated with the matter field following the definition of hair \cite{Nunez:1996xv}. We assume that $p$ and $\rho$ go to zero faster than $r^{-d}$ to satisfy this condition. Therefore,
    \begin{equation}
        P(r)\rightarrow0\,\,\,\,\,\, \text{as}\,\,\,\,\,\,\,\,r\rightarrow\infty~.
    \end{equation}
\end{enumerate}
Now, we shall study the behavior of $P(r)$ and $P'(r)$ at the horizon. From Eq. (\ref{lambdarh}) and Eq. (\ref{p}) we find that, $p(r_H)<0$. Therefore,
\begin{equation}\label{func}
    P(r_H)<0~.
\end{equation}
From Eq. (\ref{Pprime}), it follows that the second term on the right-hand side diverges. The requirement of the regularity of the horizon ensures that the left-hand side is finite on the horizon. This gives the following conditions,
\begin{equation}\label{pplusrho}
    \rho(r_H)=-p(r_H)~.
\end{equation}
Substituting Eq. \ref{lambdarh}, Eq. (\ref{p}), and Eq. (\ref{pplusrho}) into Eq. (\ref{Pprime}), we obtain
\begin{equation}\label{funcder}
    P'(r_H)<0~.
\end{equation}
Since gravity is kinematic in critical dimension $d = 2N + 1$, we have considered dimensions $d > 2N + 1$ \cite{Dadhich:2012cv}.
Next, we shall find out the radius of the photon orbit and examine the nature of $P(r)$ there. Photon orbit is the bound below which no test particle can orbit the black hole. For the line element Eq. (\ref{metric}), one finds the Lagrangian describes the geodesic in the equatorial plane ($\theta=\pi/2$) as follows
\begin{equation}\label{lagrange}
    2L=-e^\nu\Dot{t}^2+e^\lambda \Dot{r}^2+r^2\Dot{\phi}^2~,
\end{equation}
where dots represent the derivative with respect to the proper time. Corresponding conjugate momenta are given by,
\begin{equation}\label{pt}
    p_t=-e^\nu\Dot{t}\equiv-\mathcal{E}=\text{const}~,
\end{equation}
\begin{equation}\label{pphi}
    p_\phi=r^2\Dot{\phi}\equiv\mathcal{L}=\text{const}~,
\end{equation}
and 
\begin{equation}
    p_r=e^\lambda \Dot{r}~.
\end{equation}

\begin{figure}[ht]
\includegraphics[width=6cm]{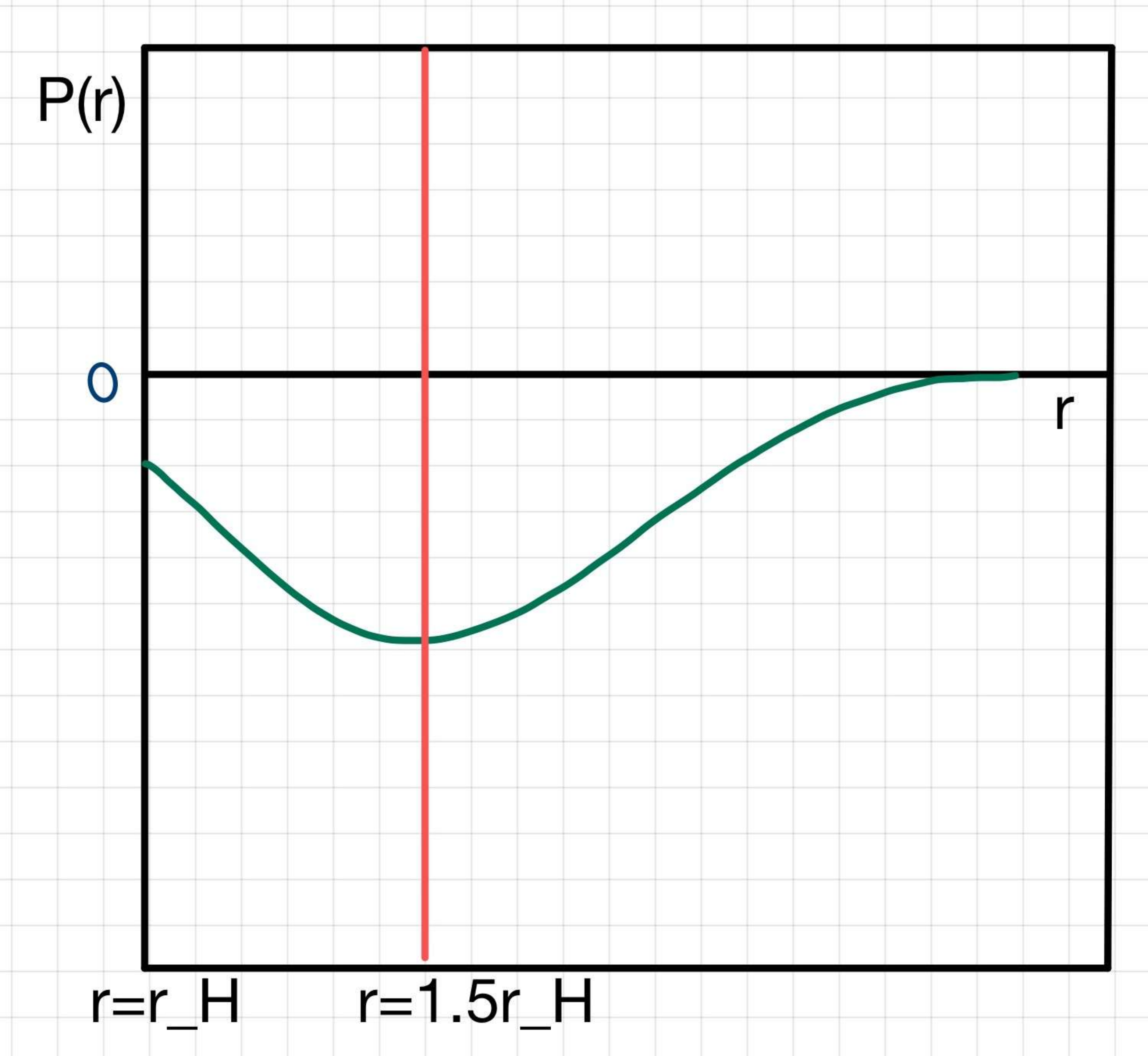}
\caption{A schematic representation depicts the behavior of the function $P(r)$ across various regions of spacetime (illustrated by the green curve), with $r$ represented on the horizontal axis. Here, $r$ denotes the coordinate distance from the center of the black hole. The vertical red line signifies the position of the local minima of $P(r)$. We illustrate the lower bound of the position of these minima. At $r=r_H$, $P(r)$ is negative, while as $r$ approaches infinity, $P(r)$ converges to 0.}
\label{figure1} 
\end{figure}

As the exterior spacetime is static and spherically symmetric, there are two constants of motion. These are represented by $\mathcal{E}$ and $\mathcal{L}$. Using Eq. (\ref{pt}), Eq. (\ref{pphi}), and Eq. (\ref{lagrange}) the Hamiltonian of the system can be written as,
\begin{equation}\label{hamilton}
    2H=-\mathcal{E}^2e^{-\nu}+\frac{\mathcal{L}^{2}}{r^2}+e^\lambda \Dot{r}^2=\epsilon=\text{const}~,
\end{equation}
where $\epsilon=0$ and $\epsilon=-1$ represent the null and timelike geodesics, respectively. 
Rearranging Eq. (\ref{hamilton}) we obtain
\begin{equation}\label{geodesic}
    \Dot{r}^2=e^{-\lambda}\left[\mathcal{E}^2e^{-\nu}-\frac{\mathcal{L}^{2}}{r^2}+\epsilon\right]~.
\end{equation}
The circular geodesics are defined by $\Dot{r}^2=(\Dot{r}^2)'=0$ \cite{chandra}. Imposing this condition into Eq. (\ref{geodesic}), we find the following relations for timelike circular geodesic,
\begin{equation}
    \mathcal{E}^2=\frac{2e^\nu}{2-r\nu'}
\end{equation}
 and
 \begin{equation}
     \mathcal{L}^2=\frac{r^3\nu'}{2-r\nu'}~.
 \end{equation}
To ensure the positivity of energy, we find the following inequality,
\begin{equation}\label{timelike}
    2-r\nu'>0~.
\end{equation}
The radius of the null circular orbit can be found from Eq. (\ref{geodesic}), which is given by
\begin{equation}\label{null}
    r_p=\frac{2}{\nu'}~.
\end{equation}
Using Eq. (\ref{p}), Eq. (\ref{timelike}) and Eq. (\ref{null}) we obtain the following condition
\begin{equation}\label{bound}
    2\mathcal{A}Ne^{-\lambda}-8\pi p-\mathcal{A}(d-2N-1)(1-e^{-\lambda})\geq0~,
\end{equation}
where $\mathcal{A}=\frac{(1-e^{-\lambda})^{N-1}}{2^{N-1}r^{2N}}$. The equality sign refers to the null circular orbits. This bound specifies the spacetime region permissible for the corresponding timelike and null geodesics. To find the characteristic of $P(r)$ in this region, we put Eq. (\ref{bound}) into Eq. (\ref{Pprime}) and use the trace condition Eq. (\ref{trace}), we obtain the following relation
\begin{equation}\label{photon}
    P'(r_p)=r_p^{d-1}T\leq0~,
\end{equation}
where $T=[-\rho+(d-1)p]$ is the trace of energy-momentum tensor. 

\textit{Analysis:} Let's analyze the functional behavior of $P(r)$ in different regions of spacetime. From the above derivation, we see that the function $P(r)$ and its derivative (Eq. (\ref{func}), Eq. (\ref{funcder})) is negative at the event horizon. Next, we find that the gradient of $P(r)$ either vanishes or is negative at the photon sphere (Eq. (\ref{photon})). According to the conditions on the energy-momentum tensor, the function $P(r)$ must vanish at asymptotic infinity. Therefore, it is clear that $P(r)$ admits a local minimum at the photon sphere ($r=r_p$) or beyond it (see Fig. (\ref{figure1})). This analysis concludes that the hair must present at least up to the photon orbit of the corresponding black hole spacetime, i.e., $r_{hair}\geq r_p$. Therefore, the matter field's nontrivial (nonasymptotic) behavior must be present at least up to the null circular geodesic starting from the horizon. Therefore, the lower bound of the length of the hair of a hairy black hole in pure Lovelock gravity is the radius of the light ring ($r_p$), independent of the dimensionality and Lovelock order.



\section{Discussion}\label{discussion}

In this paper, we've confirmed the 'no short hair conjecture' validity for black holes in pure Lovelock gravity. Through thorough examination, we've established that the hair surrounding these black holes extends at least to the photon sphere. What's intriguing is that regardless of the dimensionality or Lovelock order, the length of this hair remains consistent, highlighting a universal aspect of these gravitational phenomena.

Interestingly, our analysis doesn't require explicit knowledge of the metric coefficients. The spherical symmetry and staticity of the spacetime and strong energy condition are sufficient for our investigation. This analysis provides a roadmap for detecting the presence of hair in astrophysical measurements of pure Lovelock black holes.

A growing interest has recently been in studying hairy black holes in various systems. It would be particularly fascinating to discover such hairy black holes in pure Lovelock gravity and explicitly explore their hair's characteristics.

Furthermore, investigating the conjecture for pure Lovelock black holes while relaxing the staticity and spherical symmetry conditions presents an intriguing avenue for future research.


\appendix

\begin{acknowledgments}
We thank Chiranjeeb Sinha for proposing the problem and offering valuable insights. We are also grateful to Amit Ghosh and Pritam Nanda for their support. Finally, we thank the anonymous referee for their constructive comments, which enhanced the quality of this work.

\end{acknowledgments}

\color{black}


\end{document}